# Trends and Performance Visualization of Clutch Time in Japan's Professional B. League

Shota Shiiku[1][0009-0008-6602-5637] and Jun Ichikawa[1][0000-0002-3477-0594]

[1] Shizuoka University
`shiiku.shota.19@shizuoka.ac.jp`

**Abstract.** Reflecting the recent rise in Japanese basketball's competitiveness and pivotal outcomes in international tournaments, clutch-time performance has become increasingly critical. We examine clutch-time performance in Japan's B.League using play-by-play and box-score data from the 2023–24 season. Defining clutch time as the final five minutes with a score margin within five points, we compare efficiency and shot selection between clutch and non-clutch windows. We focus on effective field-goal percentage (eFG%) by spatial category, including In the Paint, Mid-Range, and 3-Point, and investigate league-wide and team-level trends. We further profile player archetypes in clutch minutes via K-means clustering on filtered box-score features. Results show that paint touches and clean three-point looks remain comparatively efficient under pressure, whereas mid-range outcomes are more context-dependent. Championship games exhibit tighter defense and scouting, moderating three-point accuracy for many teams. Our findings highlight the need for tactical flexibility, balancing efficient shot selection with adaptive counters to defensive pressure. We discuss strategic implications for late-game play-calling, spacing, and lineup choices, and outline limitations and future directions.

**Keywords:** Clutch time, B.League, Basketball analytics, Efficient field-goal percentage, Shot selection

## 1    Introduction

In basketball, performance by players and teams during close-end-game situations, so-called clutch time, is considered a decisive factor that determines outcomes. Clutch time refers to game states within the final five minutes when the score margin is five points or fewer [1]. Attention is drawn to performance in this phase because it directly affects the result. In recent years, interest in basketball has risen rapidly in Japan. At the Paris 2024 Olympic Games, for example, the Japanese national team contested several tight finishes. In one notable match, Japan went to overtime against host nation France and led 84–80 with 16.4 seconds remaining in regulation, but ultimately fell 94–90. Point guard Yuki Kawamura scored 29 points in that contest, yet differences in late-game strategy and experience appear to have decided the outcome, suggesting that clutch-time decisions shaped the result [2]. In the FIBA World Cup, Japan also defeated Finland 98–88 after a decisive fourth-quarter stretch sparked by Kawamura [3]. These cases suggest that, as the competitive level of Japanese basketball rises, the tactical and performance aspects of clutch time are increasingly important. This study analyzes clutch time in the B.League to address a gap in prior work that has focused predominantly on NBA data. Using play-by-play analysis, we go beyond single-number metrics such as raw shot percentages to provide richer insight, examine longer-term trends, and derive strategy implications tailored to Japanese basketball. These findings may also be effective in countries where basketball is still developing. Shot-selection theory further posits that end-of-game possession value increases when teams privilege high-efficiency looks and suppress low-yield contingencies under time pressure [4].

Clutch-time research has advanced in recent years. Sarlis et al. introduced an Estimation of Clutch Capabilities (EoCC) metric to quantify player impact during the closing minutes of NBA games, thereby evaluating influence near the end of contests [5]. Such analyses connect to assessments of efficiency and resilience under pressure, crucial in the moments that swing results. Oliver, using NBA Finals data, detailed methods to quantify complex elements of team strategy and player contribution [6]. Esteves and colleagues examined how scheduling and rest affect NBA performance [7], implying that stamina and focus in late stages influence clutch outcomes. Beyond NBA media definitions, the academic literature treats "clutch" as a distinct high-pressure performance state characterized by reliable execution under situational constraints, and systematically reviews its behavioral and contextual determinants [8]. Empirical tests of reputational "clutch" effects find mixed or modest evidence when benchmarked against teammates and baselines [9], while several quantitative theses re-examine late-game shooting performance with modern data and definitions [10,11]. Building on this literature, we conduct a quantitative play-by-play analysis centered on clutch time, characterizing league-wide and team-level patterns in Japan's B.League.

## 2     Methods

This study uses box scores and play-by-play data for all 739 games in the B.League B1 2023–24 regular season and the B.LEAGUE CHAMPIONSHIP 2023–24.

We split the play-by-play into clutch-time and non-clutch segments and compared shot success across periods. Shot locations were grouped into three areas, including In the Paint, Mid-Range, and 3-Point Area, as shown in Table 1. Beyond raw accuracy, spatial allocative efficiency links shot value to lineup-specific spacing and role distribution, providing a quantitative rationale for our area-based metrics [12]. We analyzed each area using Effective Field Goal Percentage (eFG%), computed by Eq. (1). Unlike raw FG%, eFG% reflects the 1.5× value of made three-pointers and thus better captures modern shot value [13].

$$\text{eFG\%} = \frac{\text{FGM} + 0.5 \times \text{3PM}}{\text{FGA}} \qquad (1)$$

Here, FGM is total made field goals, 3 PM is made three-pointers, and FGA is field-goal attempts. To characterize player roles during clutch time, we also performed K-means clustering on players' aggregate clutch-time statistics (points, rebounds, assists, etc.), after filtering for minimum attempt thresholds.

**Table 1.** Shot-area taxonomy.

| Category | Area |
|---|---|
| In the Paint | Under basket, in the paint |
| Mid-Range | Inside rightwing, inside right, inside center, inside left, inside leftwing |
| 3-Point Area | Outside rightwing, outside right, outside center, outside left, outside leftwing, backcourt |

## 3 Results and Discussion

Figure 1 shows eFG% by shot area during clutch and non-clutch time for the top 10 teams (Championship games are omitted here due to smaller sample sizes). Overall, eFG% in the paint and from three tends to exceed 50%, whereas mid-range is lower. This likely reflects a strategy that prioritizes higher-value or higher-probability shots. Because mid-range attempts yield only two points regardless of distance and often occur amid tighter defense, success rates are lower. Three-point eFG% remains comparatively high in both clutch and non-clutch windows, though there is a slight dip in clutch time, which is consistent with increased defensive pressure. Even so, given the value of a made three, the expected value can still be favorable in decisive moments. Consistent with data-driven clutch models, inside-out creation that yields uncontested catch-and-shoot threes outperforms static isolations, whereas low-EV contested pull-ups should be constrained to designated specialists [14]. Across all teams, we conducted $t$-tests comparing success rates between clutch and non-clutch periods. Only the three-point area showed a significant decline in clutch time ($t(46) = -3.004$, $p = 0.004$), suggesting that while top teams generally maintain stable three-point accuracy throughout the game, they still experience pressure-related declines in late-game situations.

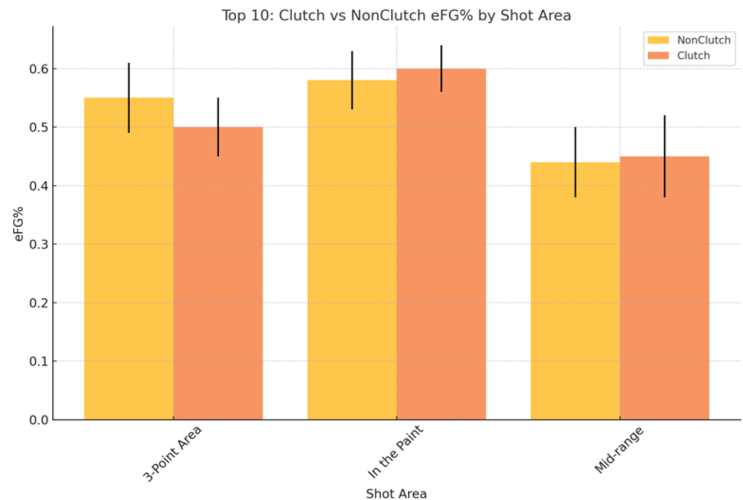

**Fig. 1.** Top-10 teams: Clutch vs Non-Clutch eFG% by shot area.

Figure 2 plots correlations between eFG% by area during clutch time and win percentage. Although *p*-values slightly exceed the 5% threshold and thus warrant cautious interpretation, three-point efficiency in clutch time shows a weak positive correlation with win percentage ($r = 0.28$, $p = 0.18$). In contrast, paint efficiency shows virtually no association ($r = 0.01$, $p = 0.98$), and mid-range efficiency exhibits a low, non-significant relationship ($r = 0.23$, $p = 0.29$). Considering base rates, mid-range shots appear less frequently attempted in clutch situations and are less determinative of outcomes. When extended to non-clutch periods, however, three-point efficiency demonstrates a moderate and statistically significant association with winning ($r = 0.63$, $p = 0.006$). This suggests that consistent three-point shooting across the full game contributes strongly to overall success, whereas late-game overreliance on deep shots may not be strictly necessary.

Figure 3 shows the league-wide change in made and attempted three-pointers between clutch and non-clutch windows. Figure 4 repeats this differential analysis for Championship teams, plotting regular-season (blue) and Championship (red) values. Teams without clutch-time possessions in the Championship have no red markers. The first quadrant in Fig. 4 indicates improvements in both three-point volume and accuracy in clutch; top regular-season teams such as Ryukyu Golden Kings**,** Nagoya Diamond Dolphins, and Alvark Tokyo fall here (black square frame), reflecting the broader modern emphasis on the three. However, most teams' accuracy declined in Championship clutch situations. For example, Chiba Jets and Nagoya Diamond Dolphins reduced attempt rates and saw accuracy drop, suggesting opponent scouting and defensive adjustments constrained their three-point strategies. Evidence on end-of-game play-type effectiveness supports denying pull-up threes and protecting the restricted area on defense while leveraging cooperative actions (e.g., ghost screens, drive-and-kick) to generate high-quality attempts on offense [15]. In contrast, a team like Ryukyu Golden Kings remained stable in both rate and accuracy, indicating consistency of strategy even under heightened stakes.

Table 2 reports K-means clusters of players based on aggregated clutch-time stats after filtering for minimum attempts ($\geq 10$ FGA and $\geq 10$ 3PA). Fig. 5 visualizes each cluster's summed stats alongside team win percentage (stacked bars for stats; blue line for win%). The clusters and their characteristics are:

- Balanced Contributor (Cluster 0):
  Average FG% and 3P%; relatively balanced contributions in rebounds and assists.
- Perimeter Shooters (Cluster 1):
  High 3P%; proficient from outside; fewer rebounds/assists; efficient scorers.
- Role Player (Cluster 2):
  Lower shooting and scoring; relatively greater emphasis on rebounding/defense.
- Star Scorer (Cluster 3):
  High FG%; strong across points, rebounds, and blocks; highest contribution to winning among clusters.

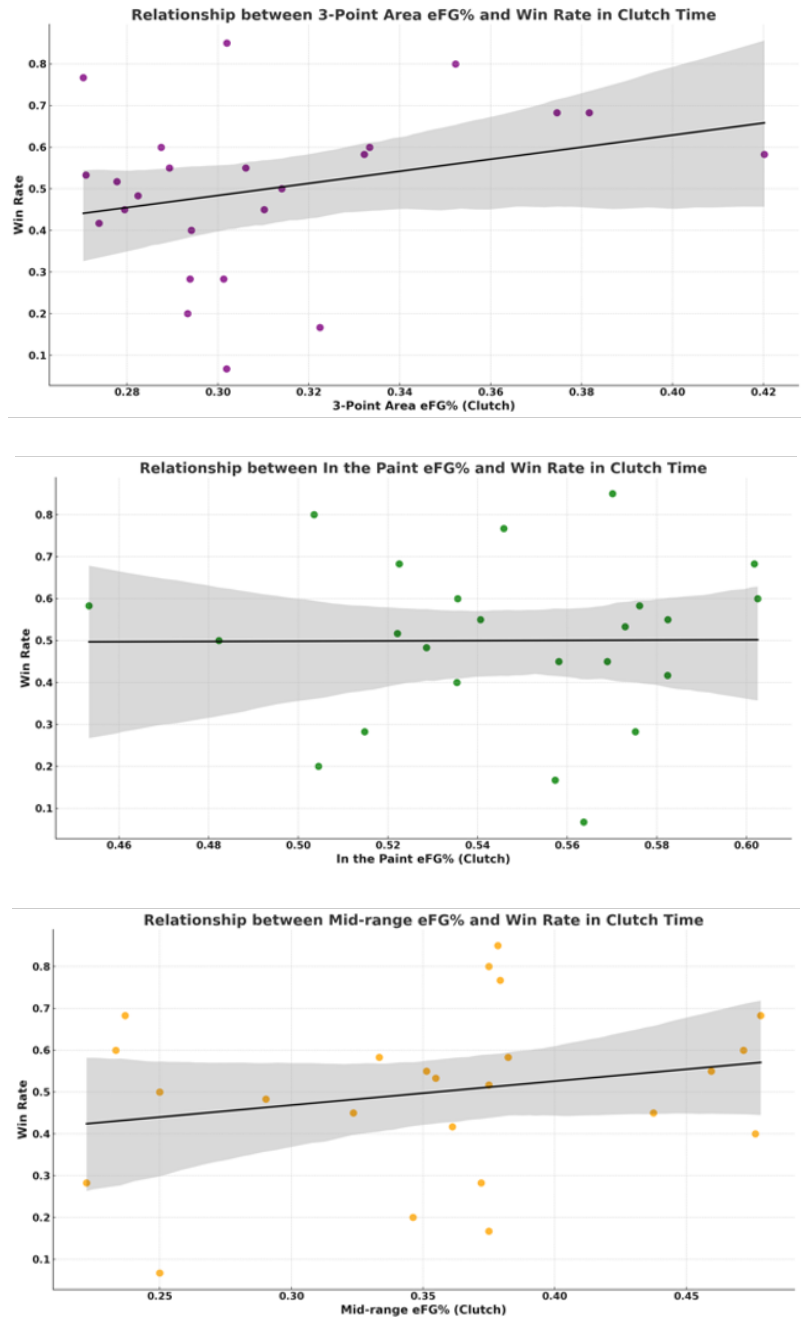

**Fig. 2.** Correlation between eFG% by shooting area during clutch time and win percentage. The black lines represent the regressions, and the gray areas indicate the confidence interval.

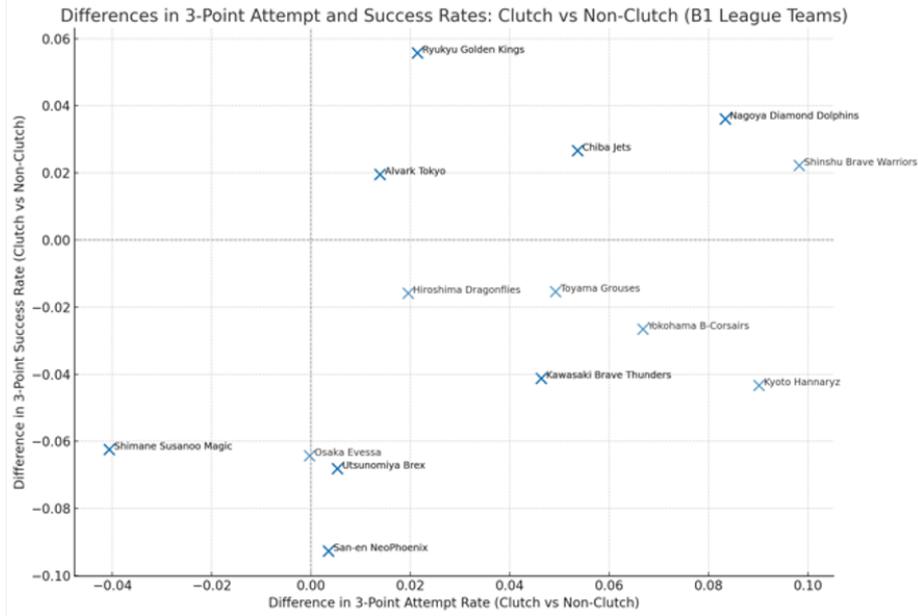

**Fig. 3.** Differences in 3-point attempt and success rates: Clutch vs Non-Clutch across the teams.

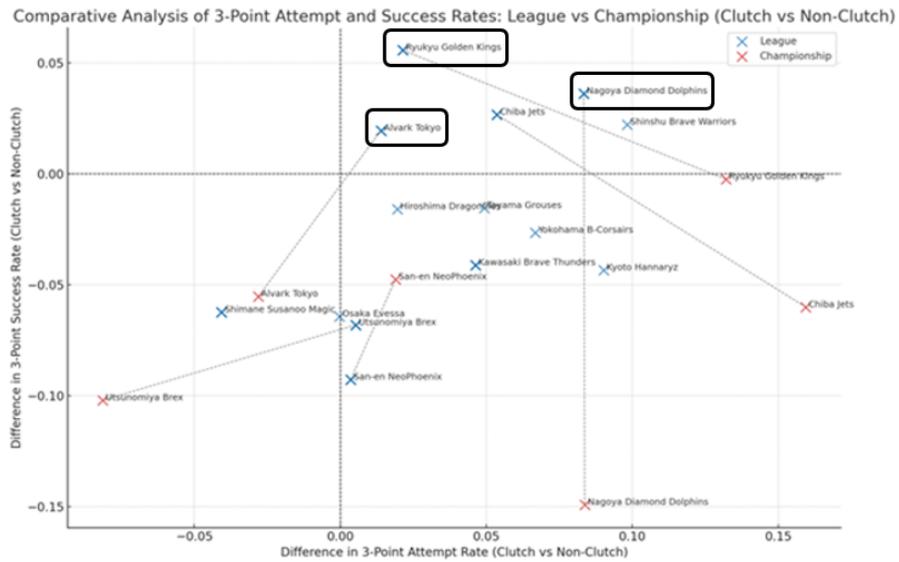

**Fig. 4.** League vs Championship comparison in 3-point attempt/success differences between Clutch vs Non-Clutch. The black square frames indicate the teams reported in the Results and Discussion section.

**Table 2.** Summary statistics by cluster (clutch time).

| Cluster | FG% | 3P% | Assists | Turnovers | Rebounds | Points | Steals | Blocks | Win% |
|---|---|---|---|---|---|---|---|---|---|
| 0 | 41.69 | 31.63 | 92.69 | 49.79 | 98.19 | 260.60 | 28.65 | 4.95 | 51.05 |
| 1 | 40.94 | 34.65 | 28.22 | 17.50 | 49.12 | 119.64 | 10.50 | 2.92 | 49.55 |
| 2 | 30.61 | 21.87 | 18.65 | 12.46 | 27.49 | 48.60 | 7.28 | 1.71 | 42.38 |
| 3 | 50.39 | 31.90 | 68.19 | 46.78 | 220.31 | 327.14 | 28.47 | 23.14 | 54.86 |

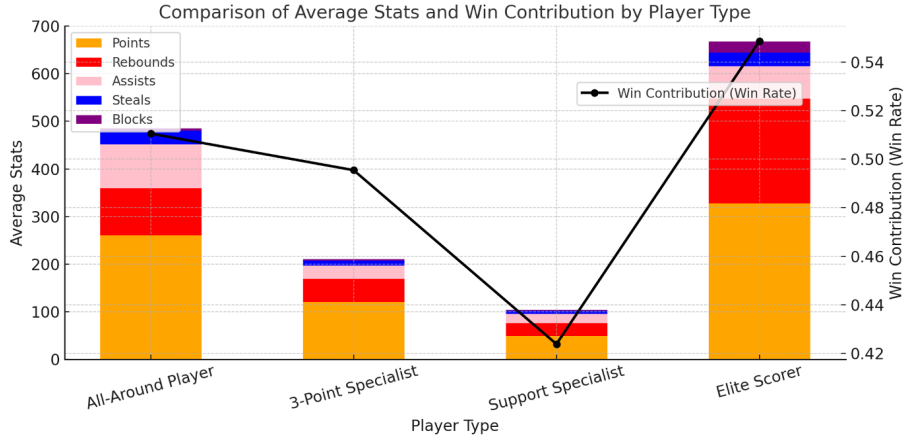

**Fig. 5.** Comparison of average stats and win contribution by player type.

## 4    Proposal, Limitations, and Future Work

Our findings offer new insight into clutch-time tactics and player performance in the B.League, although the provided dataset is limited. From an optimization perspective, end-game policy should align play-calling with value-maximizing decision rules that respect possession dynamics and clock constraints [16]. In particular, eFG% in the 3-Point Area and Paint is most closely tied to winning, consistent with the efficiency-driven shot-selection trend in modern basketball. While clutch-time three-point percentage matters, many Championship teams experienced accuracy drops, likely due to targeted opponent preparation. This underscores the need for adaptive tactics as stakes rise, such as adjusting shooter deployment and defensive schemes to preserve efficiency. Although mid-range efficiency is generally lower, it does not imply that the mid-range is always an inferior choice. Depending on opponent coverage and situational constraints, mid-range jumpers can be effective counters. Figure 4 suggests that teams with rigid clutch-time shot profiles are more vulnerable to performance declines; precisely because the three is emphasized, diversifying options can be valuable.

We defined clutch time as the final five minutes within five points, but pivotal moments exist outside this window. For instance, bursts that shrink or extend leads can

8      S. Shiiku and J. Ichikawa8      S. Shiiku and J. Ichikawa

be decisive. Expanding the analysis to such phases may offset the relative sparsity of clutch events while testing tactical efficacy. Within the clutch, further stratifying by score margin (e.g., 1-point vs.≥ 2-point games) could illuminate how optimal strategies and rotations shift with leverage. For the clustering analysis, incorporating richer context (e.g., game flow, score differential, foul trouble) beyond basic box metrics could yield more actionable insights. Identifying players who excel not only in clutch but also during *negative* momentum phases could inform novel deployment patterns. Future work will pursue such temporal segmentation and additional features to deepen strategic understanding of clutch situations in the B.League.

**Acknowledgments.** This work was supported by the Sports Data Science Subcommittee of the Japan Statistical Society, School of Statistical Thinking at the Institute of Statistical Mathematics of the Research Organization of Information and Systems, and Data Stadium, Inc. We gratefully acknowledge their assistance.